\documentstyle[11pt,amssym,emulateapj_rj,epsfig]{article}

\newcommand{\ie}{\mbox{i.e.}}%
\newcommand{\eg}{\mbox{e.g.}}%
\newcommand{\etal}{\mbox{et al.\ }}%
\newcommand{\MB}{\mbox{$M_B$}}%
\newcommand{\BR}{\mbox{$(B\!-\!R)$}}%
\newcommand{\BRe}{\mbox{\BR$_e$}}%
\newcommand{\UB}{\mbox{$(U\!-\!B)$}}%
\newcommand{\UBe}{\mbox{\UB$_e$}}%
\newcommand{\EBV}{\mbox{E(\bv)}}%
\newcommand{\tpm}{\mbox{$\pm$}}%
\newcommand{\tsim}{\mbox{$\sim$}}%
\newcommand{\HI}{\mbox{H\,{\sc i}}}%
\newcommand{\HII}{\mbox{H\,{\sc ii}}}%
\newcommand{\Ha}{\mbox{H$\alpha $}}%
\newcommand{\Hb}{\mbox{H$\beta $}}%
\newcommand{\OII}{\mbox{[O\,{\sc ii}]}}%
\newcommand{\OIII}{\mbox{[O\,{\sc iii}]}}%
\newcommand{\lam}{\mbox{$\lambda $}}%
\newcommand{\Rindex}{\mbox{R}$_{23}$}

\topline{The Astrophysical Journal, 000:000--000, 2001
        \hspace*{95mm} (MS-52562-ApJ)}
\submitted{Received 2000 August 25; accepted 2000 December 22}
\received{August 25, 2000}
\accepted{December 22, 2000}
\journalid{000}{2001 \null }
\articleid{000}{000}
\paperid{MS-52652-ApJ}
\lefthead{JANSEN, FRANX, \& FABRICANT}
\righthead{[O\,{\small II}] AS A TRACER OF STAR FORMATION}

\begin{document}

\title {\Large\bf [O\,{\large\bf II}] as a tracer of current star formation}

\author{Rolf A. Jansen,$^{1,2,3}$ Marijn Franx,$^{4}$ \&
	Daniel Fabricant$^{2}$}

\affil {$^1$Kapteyn Astronomical Institute, Postbus~800, NL-9700~AV
        Groningen, The Netherlands}
\affil {$^2$Harvard-Smithsonian Center for Astrophysics, 60 Garden St.,
        Cambridge, MA~02138}
\affil {$^3$Astrophysics Division, Space Science Department of ESA,
	ESTEC, Postbus~299,\\ NL-2200 AG Noordwijk, The Netherlands}
\affil {$^4$Leiden Observatory, Postbus 9513, NL-2300~RA Leiden,
        The Netherlands}
\affil {rjansen@astro.estec.esa.nl, franx@strw.leidenuniv.nl,
	dfabricant@cfa.harvard.edu}


\begin{abstract}

\OII\lam3727\AA\ is often used as a tracer of star formation at
intermediate redshifts ($z\gtrsim0.4$), where \Ha\ is not easily
observed.  We use the spectrophotometric data of the Nearby Field Galaxy
Survey to investigate the range and systematic variation in observed
\OII/\Ha\ emission line ratio as a function of galaxy luminosity at low
redshift.  We find that the observed \OII/\Ha\ ratio varies by a factor
of seven at luminosities near \MB$^{\!*}$.  The \OII/\Ha\ ratio is
inversely correlated with luminosity.  The scatter in the \OII/\Ha\
ratio and the dependence of the ratio on luminosity are due in equal
parts to reddening and to the metallicity dependent excitation of the
ISM.  The uncertainty in star formation rates derived from \OII\ fluxes
is therefore large.  If \Ha\ cannot be observed, high S/N \Hb\ fluxes
are much preferable to \OII\ fluxes for deriving star formation rates. 
We present several purely empirical corrections for extinction. 

\end{abstract}


  
\section{Introduction}
 
The \OII\lam3727\AA\ doublet is by far the strongest emission line in
the blue and is readily observable even in low S/N spectra.  Studies by
Gallagher, Bushouse, \& Hunter (1989) and Kennicutt (1992) showed that
the line strengths of \OII\lam3727 and the primary star formation
tracers \Hb\ and \Ha\ are correlated, offering a convenient means of
measuring star formation rates (SFRs) where \Ha\ has shifted out of the
optical regime ($z\gtrsim0.5$).  However, \OII\ is much less directly
coupled to the intensity of the UV radiation from hot, young stars in
\HII\ regions than the \Ha\ recombination emission, and can be expected
to be a noisier measure of the SFR.  In particular, the \OII\ emission
depends on the elemental abundances and ionization state of the parent
\HII\ region.  In addition, the (typically uncertain) correction for
extinction is more important at \OII\lam3727 than at \Ha.  Nonetheless,
Gallagher \etal (1989) and Kennicutt 1992 reported good correlations of
\OII\ to \Hb\ and \OII\ to \Ha\ line strengths, respectively, with a
scatter of $\sim$50\% about the mean relation.  Gallagher \etal obtained
large aperture spectrophotometry of 75 blue irregular galaxies, while
Kennicutt obtained integrated spectrophotometry of 90 normal and
peculiar galaxies.  In Gallagher et al.'s sample the ratio of \OII\ to H
recombination line strengths are higher ($\sim$50\%) on average than in
Kennicutt's sample.  Recently, Tresse \etal (1999) have confirmed the
Gallagher \etal and Kennicutt result using a sample of $\sim$1700 nearby
galaxies ($z\lesssim0.1$), finding an intermediate ratio of \OII\ to
\Ha. 

Hammer \etal (1997) compare the \OII\ and \Ha\ equivalent widths (EWs)
for a sample of 61 field galaxies at $z<0.3$ identified in the
Canada-France Redshift Survey (CFRS), and find that the \OII\ EW is   
typically about a factor of two higher than predicted by the Kennicutt
(1992) relation.  Hammer \etal suggest that the strong \OII\ emission
in their sample may be a result of lower extinction and metallicity in
the CFRS sample, which contains galaxies of lower average luminosity
than the Kennicutt (1992) sample.  Hammer \etal suggest that these
factors become still more important at high $z$.

Using the high signal to noise spectrophotometric data from the Nearby
Field Galaxy Survey (NFGS; Jansen \etal 2000b) we also found that the
\OII/\Ha\ ratio is higher for lower luminosity galaxies, confirming the
suggestion of Hammer \etal (1997).  Here, we use the NFGS data to
investigate the roles of extinction and metallicity, both of which are
correlated with luminosity, in producing the observed dependence
of the \OII/\Ha\ ratio on luminosity. 

The NFGS sample consists of 196 nearby galaxies, objectively selected
from the CfA redshift catalog (CfA~I, Huchra \etal 1983) to span a large
range of $-14$ to $-22$ in absolute $B$ magnitude without morphological
bias.  The sample should therefore be representative of the galaxy
population in the local universe, subject only to the biases inherent in
B selection and the surface brightness limit of the Zwicky catalog.  Our
data include nuclear and integrated spectrophotometry (covering the
range 3550--7250 at \tsim6\AA\ resolution), supplemented by $U$, $B$,
$R$ surface photometry from which we obtained total magnitudes and
colors.  For the details of the galaxy selection and for the $U$, $B$,
$R$ photometry, we refer the reader to Jansen \etal (2000a); for a full
description of the spectrophotometric data we refer to Jansen \etal
(2000b). 

Of the 191 normal (non-AGN dominated) galaxies in the NFGS sample, 141
show \Ha\ emission in their integrated spectra.  We will concentrate,
however, on the 118 galaxies with \Ha\ EW $\leq$$-10$\AA\ for which the
\OII/\Ha\ ratios are reliable, and on the 85 galaxies that also have
EW(\Hb)$\leq$$-5$\AA\ for which we can derive accurate Balmer decrements
to correct the observed emission line fluxes for reddening.  Throughout
the paper we correct observed emission line fluxes for Galactic
extinction, estimated from the \HI\ maps of Burstein \& Heiles (1984)
and listed in the RC3 (de~Vaucouleurs \etal 1991).  The wavelength
dependence of the extinction is assumed to follow the optical
interstellar extinction law of Nandy \etal (1975) (as tabulated in
Seaton 1979, but adopting $R_V=A_V/\EBV=3.1$ instead of 3.2).  Observed
Balmer emission line fluxes have also been corrected for stellar
absorption lines. 

We partially compensated for the stellar absorption underlying the \Hb\
emission by placing the limits of the measurement window well inside the
absorption trough, closely bracketing the emission line.  We evaluated
the residual absorption using the spectra of galaxies with no detectable
emission.  On average, an additional correction of 1 \AA\ (EW) was
required.  Similarly, for \Ha\ we found that a correction of 1.5 \AA\
was needed.  After correction, a fit to the data in a \Hb\ versus \Ha\
plot passes through the origin.  Because we restrict our analysis to
galaxies with moderate to strong emission lines, we are insensitive to
the details of this procedure. 

\vspace*{1mm}

This paper is organized as follows.  In section~2 we evaluate the range
in the observed and reddening corrected \OII/\Ha\ ratio and its
dependence on luminosity, and we demonstrate the effect of differences
in the excitation state of the interstellar medium (ISM).  In \S~3 we
relate the observed \OII\ and \Ha\ fluxes\linebreak

\newlength{\txw}
\setlength{\txw}{\textwidth}

\noindent\leavevmode
\makebox[0.475\txw]{
   \centerline{\epsfig{file=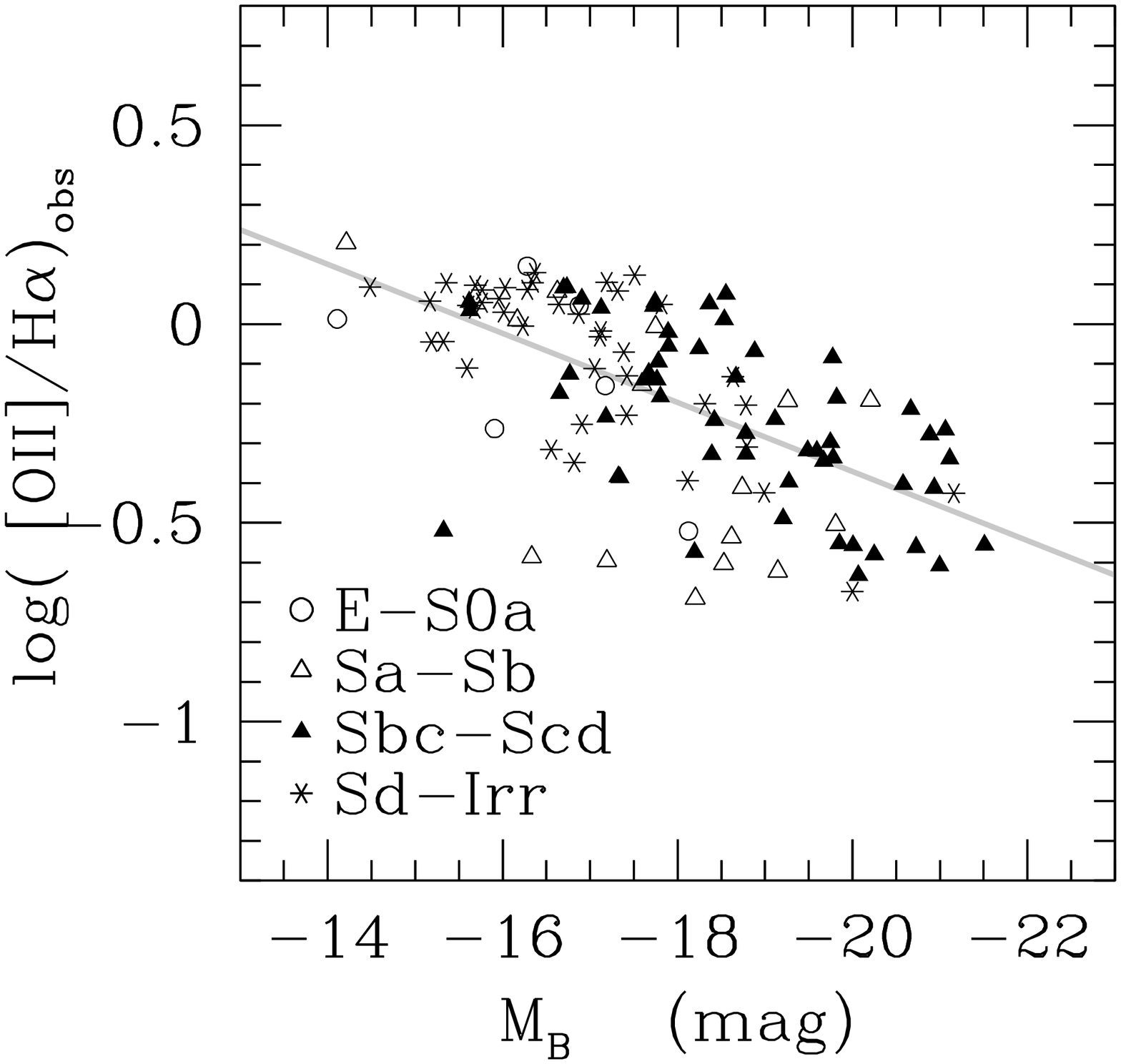,width=0.475\txw,clip=}}
}\par\vspace*{3mm}\noindent\makebox[0.475\txw]{
\centerline{
\parbox[t]{0.475\txw}{\footnotesize {\sc Fig.~1 ---} The logarithm of the
ratio of the observed \OII\ and \Ha\ emission line fluxes versus total
absolute $B$ magnitude.  Only data for the 118 galaxies with strong
emission [EW(\Ha)$<$$-10$\AA] are shown to ensure reliable emission line
ratios.  The plotting symbols are coded according to morphological type,
as indicated.  Overlayed is a linear least-squares fit to the data
points.  The \OII/\Ha\ ratio decreases systematically from \tsim1.5 at
\MB=$-14$ to \tsim0.3 at \MB=$-21$. } }
}\vfill
%

\noindent to the ionizing flux and show that these results are
consistent with previous studies.  We briefly discuss the implications
for the use of \Hb\ as SFR tracer in \S~4.  We discuss how to improve
SFR estimates and present empirical corrections in \S~5.  The
implications for evolutionary studies are discussed in \S~6.  We
conclude with a short summary (\S~7). 

\vfill

  
\section{Variation in the \OII/\Ha\ ratio}

In figure~1 we plot the ratio of the observed \OII\ and \Ha\ emission
line fluxes vs.\ total absolute $B$ magnitude for the 118 galaxies with
EW(\Ha)$<$$-10$\AA.  The Balmer emission line fluxes were corrected for
stellar absorption, but neither \Ha\ nor \OII\ are corrected for
internal interstellar reddening.  The observed \OII/\Ha\ ratio decreases
systematically with increasing galaxy luminosity, from \tsim1.5 at
\MB=$-14$ to \tsim0.3 at \MB=$-21$, but the range is large.  At
\MB=$-18$, only a magnitude fainter than the characteristic absolute
magnitude of the local luminosity function ($M_*\sim -19.2$,
de~Lapparent, Geller, \& Huchra 1989; $M_*\sim -18.8$, Marzke \etal
1994), the observed \OII/\Ha\ ratios vary by a factor of \tsim7. 

\newpage

\noindent\leavevmode
\makebox[\txw]{
   \centerline{
      \epsfig{file=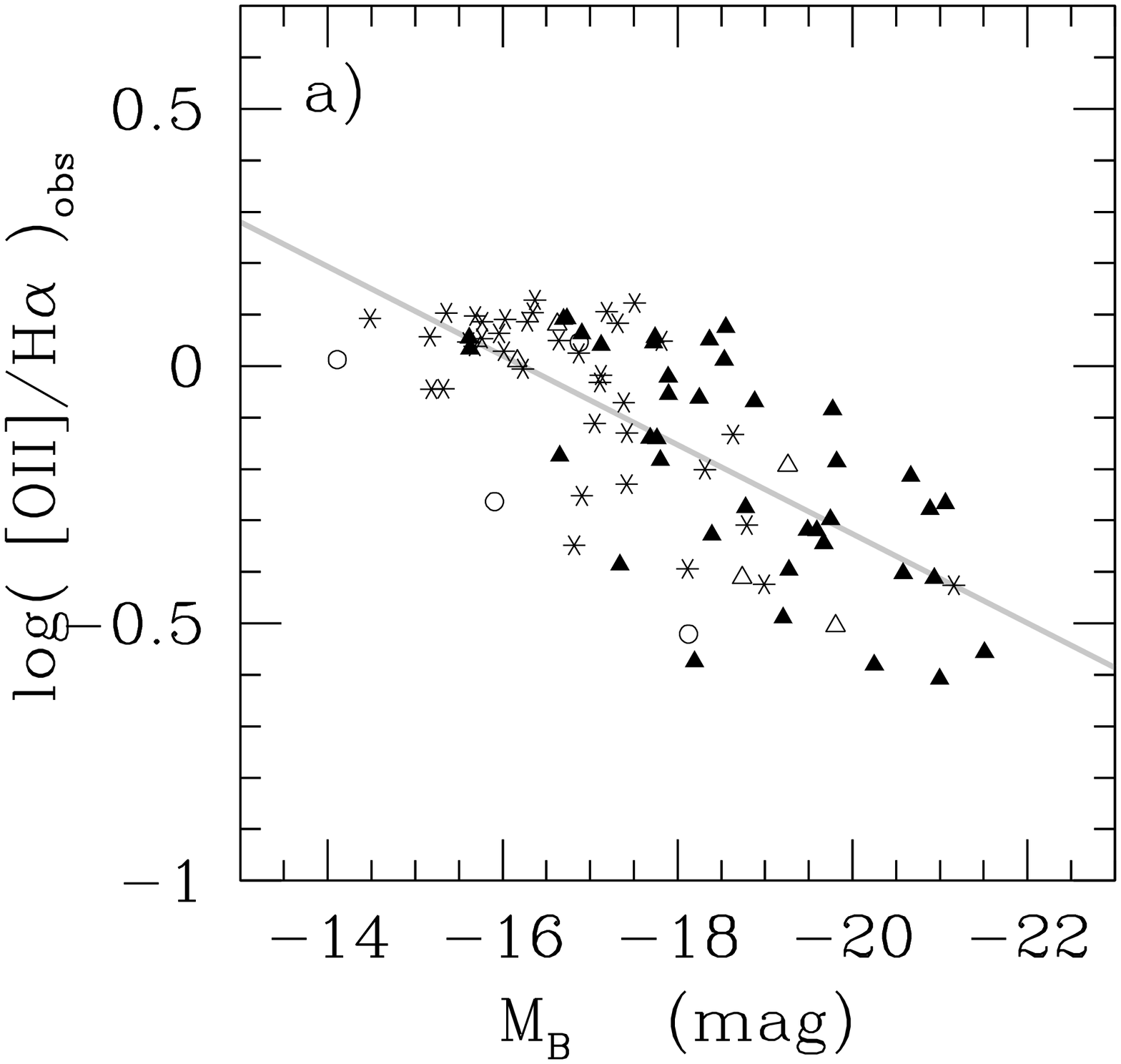,width=0.45\txw,clip=}\hspace*{0.03\txw}
      \epsfig{file=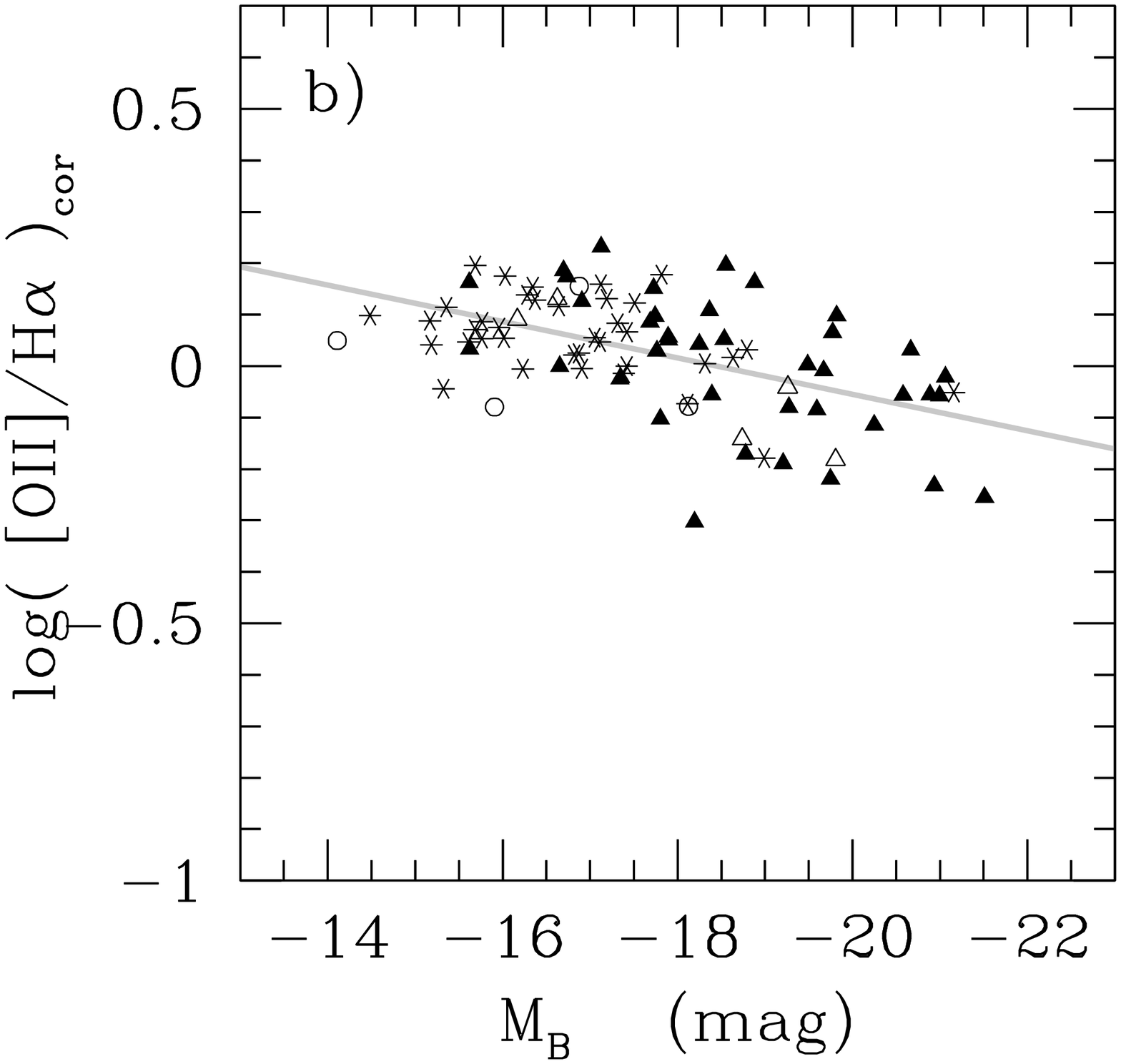,width=0.45\txw,clip=}
   }
}\par\vspace*{2mm}\noindent\leavevmode
\makebox[\txw]{
   \centerline{
      \epsfig{file=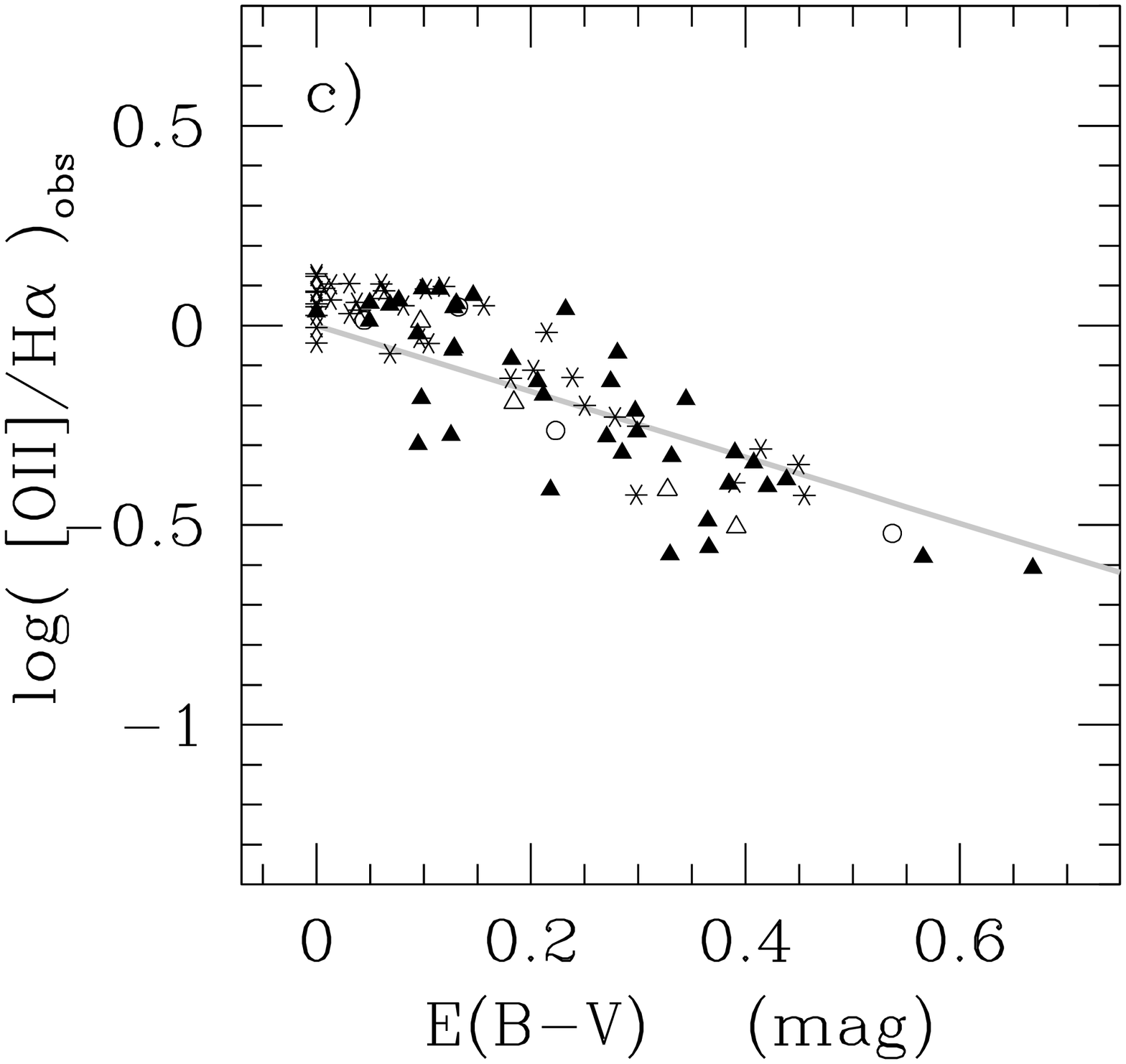,width=0.45\txw,clip=}\hspace*{0.03\txw}
      \epsfig{file=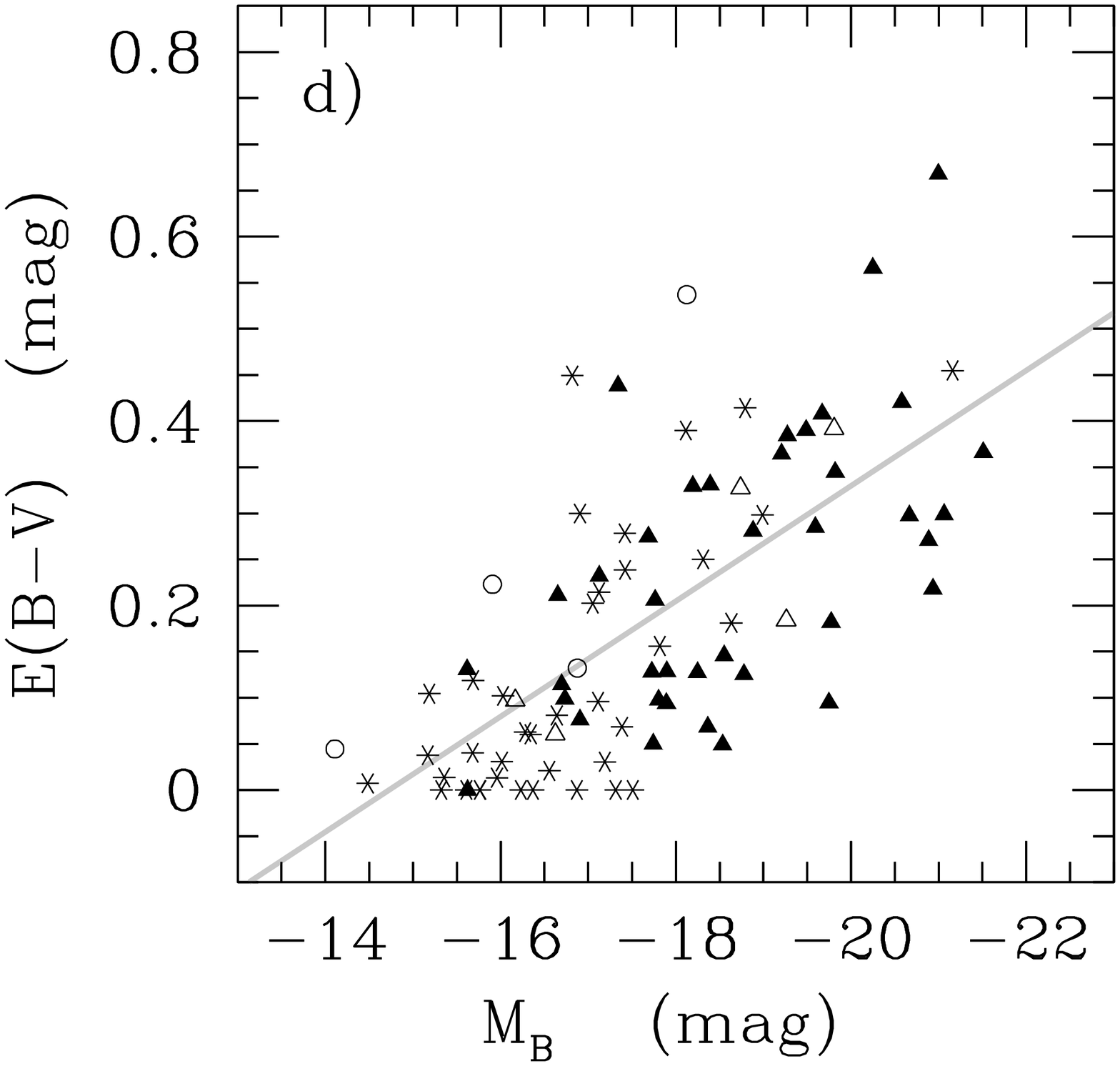,width=0.45\txw,clip=}
   }
}\par\vspace*{4mm}\noindent\leavevmode
\makebox[\txw]{
\centerline{
\parbox[t]{\txw}{\footnotesize {\sc Fig.~2 ---} (\emph{a}) Logarithm of
the observed \OII/\Ha\ ratio versus total absolute $B$ magnitude.  Here
and in the following panels we only include the 85 galaxies with
EW(\Hb)$<$$-5$\AA, allowing us to measure the Balmer decrement,
$\Ha/\Hb$, reliably.  The plotting symbols are coded according
to morphological type as in Fig.~1. (\emph{b}) Logarithm of the
\OII/\Ha\ ratio after correction for interstellar reddening versus
absolute $B$ magnitude.  The dependence on galaxy luminosity is greatly
reduced after correction for reddening, and both the total scatter and
the scatter around the best linear fit through the data has decreased.  
(\emph{c}) Logarithm of the observed \OII/\Ha\ ratio versus color excess
\EBV, determined from the observed Balmer decrement.  The \OII/\Ha\
ratio decreases as the reddening increases.  The line shows the relation
expected for the adopted reddening curve.  (\emph{d}) Color excess \EBV\
versus absolute $B$ magnitude.  Larger reddening values tend to
correspond to higher galaxy luminosities, although the range in \EBV\ at
a given galaxy luminosity is large.  A linear least-squares fit to the
data points is overlayed. } }
}\vspace*{5mm}
%

\subsection{Effects of Reddening by Dust}

To evaluate the contribution of extinction to the variation in the
\OII/\Ha\ ratio and to the ratio's dependence on galaxy luminosity, we
compute the color excess, \EBV, from the observed Balmer decrement,
$\Ha/\Hb$.  We assume an intrinsic ratio of 2.85\linebreak

\null\vspace*{1.10\txw}\noindent (the Balmer decrement for case B
recombination at T$=$$10^4$ K and $n_e\sim10^2\hbox{--}10^4$ cm$^{-3}$;
Osterbrock 1989) and adopt the optical interstellar extinction law of
Nandy \etal (1975) (as tabulated in Seaton 1979, but adopting
$R_V=A_V/\EBV=3.1$ instead of 3.2).  This procedure results in a
\emph{lower limit} to the actual reddening, as the observed Balmer
decrement will be weighted to the regions of lowest line-of-sight
extinction (Kennicutt 1998a).  Other studies suggest, however, that this
approach gives a reasonable estimate of the extinction (\eg, Calzetti,
Kinney, \& Storchi-Bergmann 1994; Buat \& Xu 1996), except for the
dustiest regions in a galaxy. 

Figure~2\emph{a} is the same as figure~1 except that we include only the
85 galaxies with EW(\Hb)$<$$-5$\AA, where we could measure the Balmer
decrement reliably.  Comparison of figures~1 and 2\emph{a} indicates
that exclusion of the galaxies with fainter \Hb\ emission is unlikely to
bias our discussion. 

In figure~2\emph{b} we present reddening corrected ratios of the \OII\
and \Ha\ emission line fluxes versus absolute $B$ magnitude on the same
scale as figure~2\emph{a}.  The dependence of \OII/\Ha\ on galaxy
luminosity decreases markedly after correction for interstellar
reddening: the slope of a linear least-squares fit to the data changes
from 0.087 dex/mag to 0.035 dex/mag and both the total scatter and the
scatter around the fit decrease from 0.18 to 0.09 dex and from 0.12 to
0.08 dex\footnote{ Throughout this paper we will express scatter in
terms of the Median Absolute Deviation (MAD), defined as $\frac{1}{N}
\sum_1^N |x_i-\langle x\rangle|$.  This statistic is similar to the rms,
but is less sensitive to outlying data values, $x_i$, and uses the
median, $\langle x\rangle$, rather than the mean.}, respectively.  This
implies that more than half of the observed scatter in \OII/\Ha\ must be
due to dust.  The total variation in \OII/\Ha\ decreases from a factor
\tsim 7 to \tsim 3. 

Ratios of (observed) equivalent widths behave much like extinction
corrected flux ratios, but their dependence on the stellar continuum
introduces a dependence on the star formation history of the galaxy.  If
we use EW ratios rather than reddening corrected flux ratios in
figure~2\emph{b} the scatter around the best fit increases (0.11 versus
0.08 dex) and the slope steepens slightly (0.043 versus 0.035 dex/mag). 

To show the effect of reddening on the relative strengths of \OII\ and
\Ha\ more directly, in figure~2\emph{c} we plot the logarithm of the
observed \OII/\Ha\ ratio versus color excess \EBV.  We find a clear
trend toward lower \OII/\Ha\ ratios in galaxies with larger interstellar
reddening.  This trend spans most of the total range in \OII/\Ha.  For
reference we overlay the relation expected for the adopted reddening
law, assuming an intrinsic \OII/\Ha\ ratio of 1 (\ie, fixing the
zeropoint to $\log\,(\OII/\Ha)=0$ for \EBV=0).  Although the data points
do not exactly follow the expected relation, the correspondence is good. 

In figure~2\emph{d} we plot color excess \EBV\ versus absolute $B$
magnitude.  More luminous galaxies tend to show more reddening than low
luminosity galaxies.  The scatter on this trend is very large, however. 

These figures show that galaxy luminosity, internal reddening and the
observed \OII/\Ha\ line ratios are linked.  The trend of \OII/\Ha\ with
reddening has the smaller scatter.  Reddening, therefore, is an
important factor in the observed trend of \OII/\Ha\ with \MB, but it is
not uniquely related to galaxy luminosity.

\subsection{Effect of Excitation and Metallicity}

The luminosity dependence of the \OII/\Ha\ ratio does not disappear
completely after correction for reddening.  Might the variation in the
reddening corrected \OII/\Ha\ ratios result from differences in the
excitation state and metallicity of the ISM? Differences in excitation
and metallicity affect the relative strengths of the \OII\, \OIII\, and
the hydrogen recombination lines. 

Figure~3\emph{a} shows the reddening corrected \OII/\Hb\ versus
\OIII\lam\lam4959,5007/\Hb\ flux ratios for the 85 galaxy subsample. 
Here, we used \Hb\ rather than \Ha\ (note that $\Ha=2.85\,\Hb$ after
reddening correction) to allow a direct comparison of the data with the
theoretical predictions of McCall, Rybski, \& Shields (1985) for the
line ratios of \HII\ regions in galaxies as a function of metallicity. 
Along the track, the metallicity is high at the lower left (low
excitation) and low at the upper right (high excitation).  The small
systematic deviations (\tsim0.1 dex) from the model are in the same
sense that McCall \etal (1985) reported for their comparison of the
model and observations.  The scatter of 0.06 dex of the data around the
model track is consistent with the errors in the data points.  The
residuals from the model track are not correlated with either absolute
$B$ magnitude, reddening \EBV, galaxy color, or \Ha\ emission line
strength. 

The tight coupling between \OII/\Ha\ and metallicity can be demonstrated
with the metallicity sensitive ratio
$\log \,\{(\OII+\OIII\lam\lam4959,5007)/\Hb$\}, commonly denoted as
\Rindex\ (Pagel \etal 1979; Edmunds \& Pagel 1984), which we use as a
quantitative indicator of the oxygen abundance (Pagel 1997; Kennicutt
\etal 2000).  In figure~3\emph{b} we plot the reddening corrected
\OII/\Ha\ ratios versus \Rindex\/.  \Rindex\/ is computed using
reddening corrected flux ratios.  The metallicity measurements are
degenerate for values of
\Rindex = $\log\,\{(\OII+\OIII)/\Hb\}\gtrsim0.75$ (metallicities lower
than \tsim0.6 times solar when adopting the expansion in \Rindex\ given
by Zaritsky, Kennicutt, \& Huchra 1994).  This is indicated in the
figure by a dotted line. 

The \OII/\Ha\ ratio is strongly correlated with metallicity.  For
\Rindex$<$0.75 (where the metallicity determination is non-degenerate) we
can approximate the dependence on metallicity with a linear relation:
\begin{displaymath}
   \log\,(\hbox{\OII/\Ha})_{\hbox{\scriptsize cor}} \; = \;
			(0.82\pm0.03)\,\hbox{\Rindex} - (0.48\pm0.02)
\end{displaymath}
The scatter around this relation is only 0.023 dex, consistent with
little or no intrinsic error. \pagebreak

\noindent\leavevmode
\makebox[\txw]{
   \centerline{
      \epsfig{file=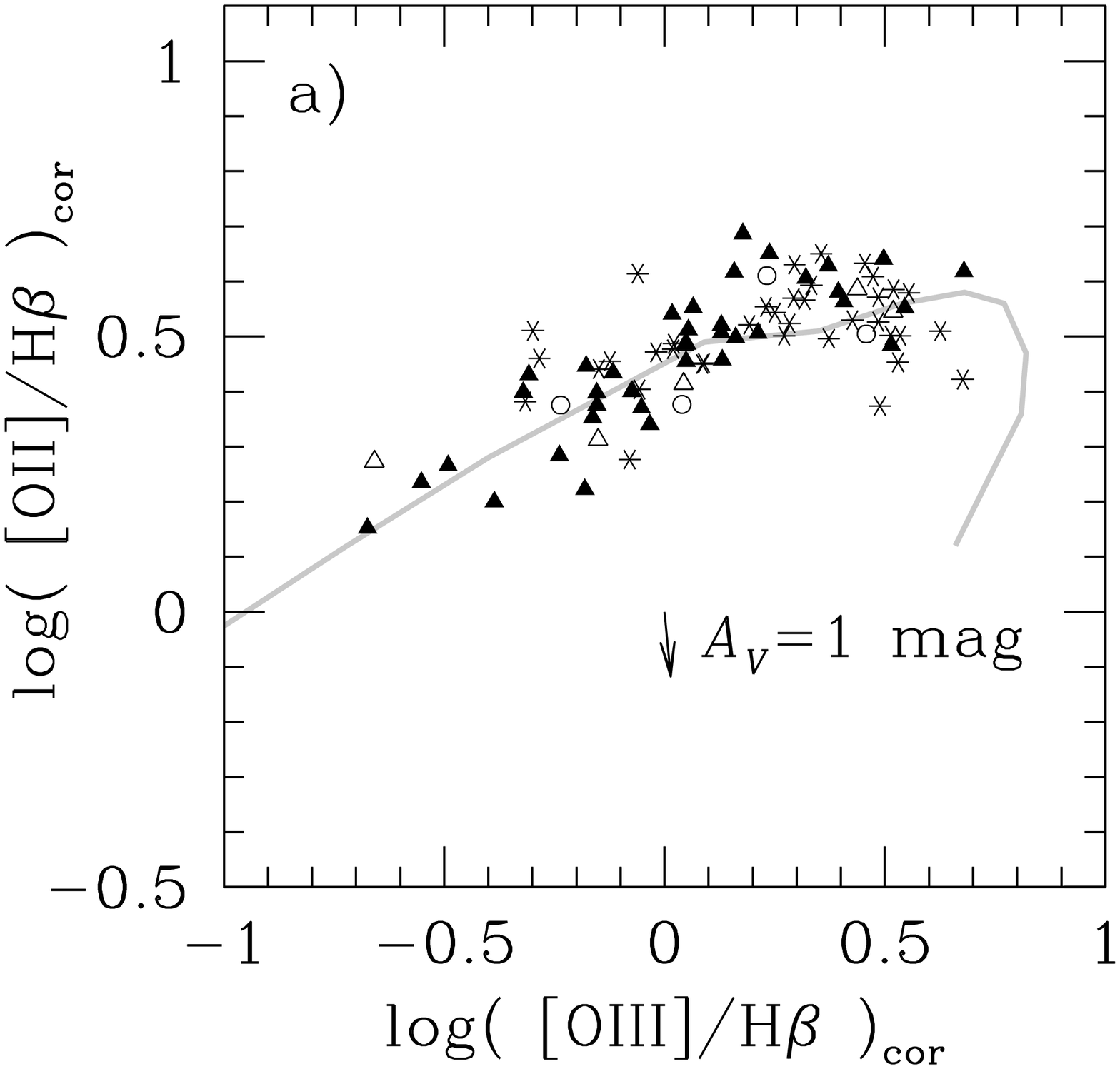,width=0.45\txw,clip=}\hspace*{0.03\txw}
      \epsfig{file=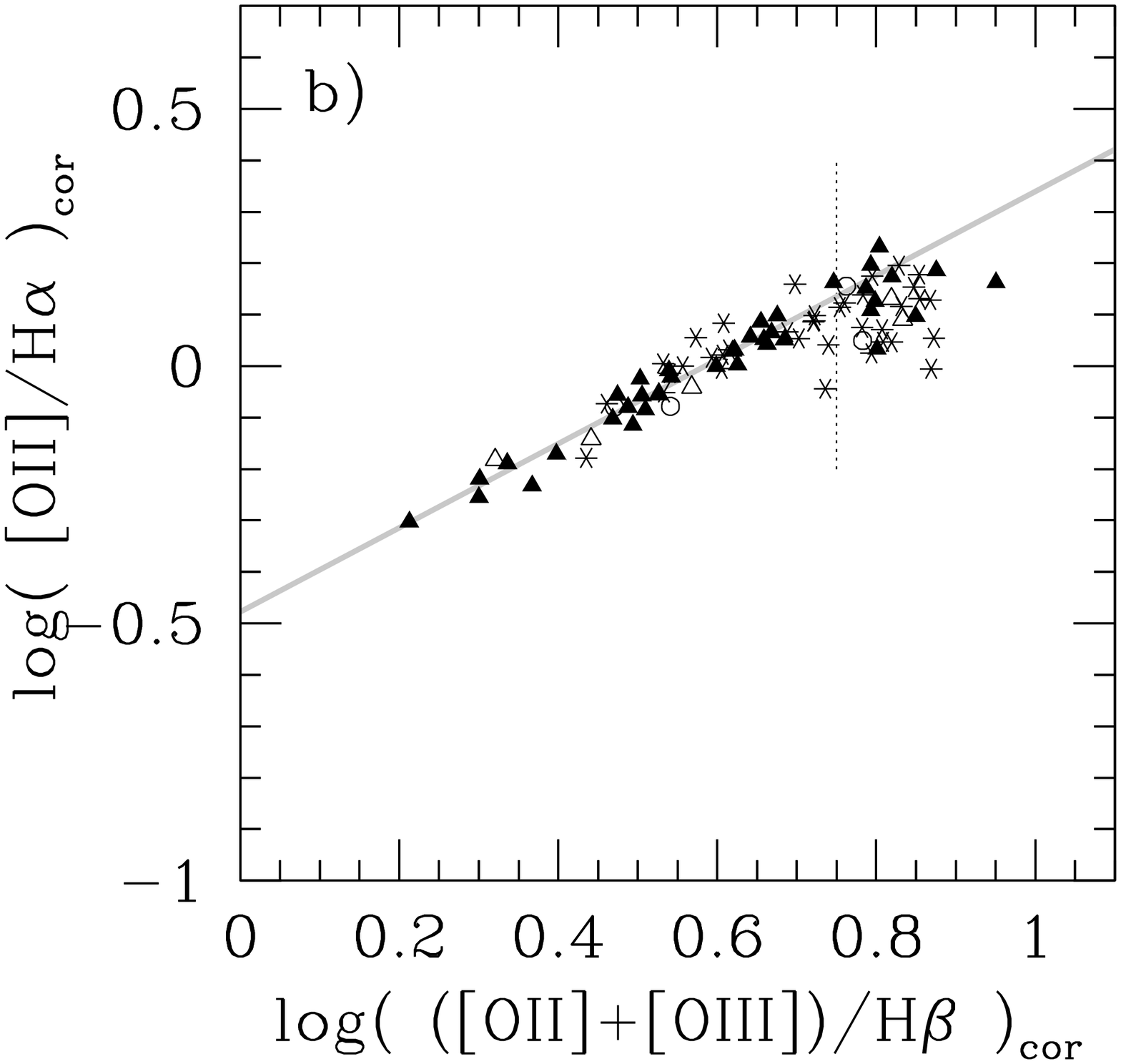,width=0.45\txw,clip=}
   }
}\par\vspace*{2mm}\noindent\leavevmode
\makebox[\txw]{
   \centerline{
      \epsfig{file=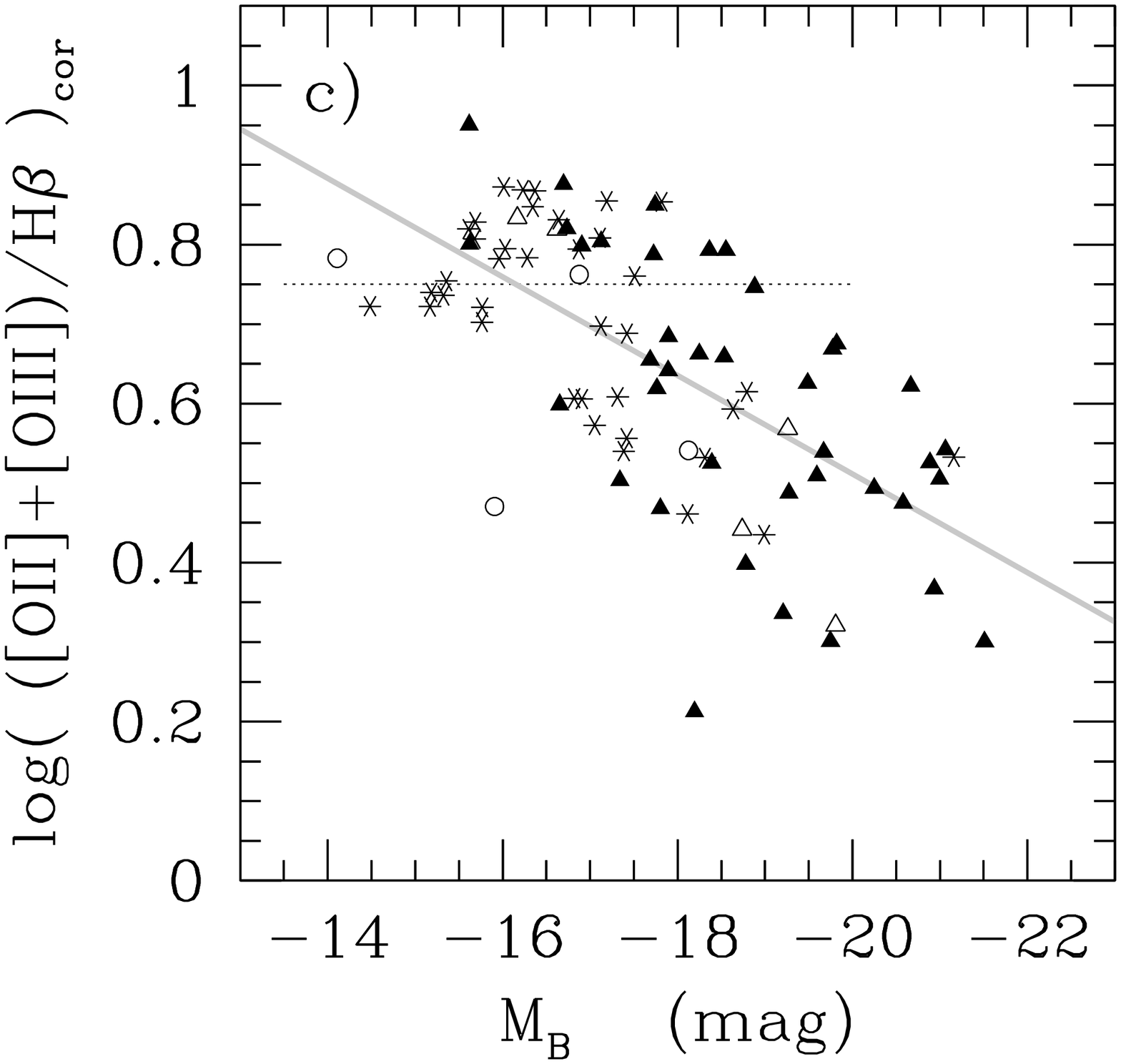,width=0.45\txw,clip=}\hspace*{0.03\txw}
      \makebox[0.45\txw]{\rule{0pt}{0.45\txw}}
   }
}\par\vspace*{4mm}\noindent\makebox[\txw]{
\centerline{
\parbox[b]{\txw}{\footnotesize {\sc Fig.~3 ---} (\emph{a}) Reddening
corrected \OII/\Hb\ versus \OIII$\lambda\lambda$4959,5007/\Hb\ flux
ratios.  The theoretical sequence of McCall \etal (1985) is overlayed. 
The scatter of the data around the model curve is consistent with the
errors in the data points.  Low excitations are found at the lower left
of the model curve, high excitations at the upper right.  (\emph{b})
Reddening corrected \OII/\Ha\ ratio versus the metallicity sensitive
index $\mbox{R}_{23} = \log\,\{(\OII + \OIII)/\Hb\}$.  The metallicity
determination is degenerate for $\mbox{R}_{23} \gtrsim 0.75$ (indicated
by the dotted line).  The \OII/\Ha\ ratio and the oxygen abundance are
strongly correlated.  The data points for $\mbox{R}_{23} < 0.75$ follow
a well defined sequence with little or no intrinsic scatter.  (\emph{c})
$\mbox{R}_{23}$ versus total absolute $B$ magnitude.  The dotted line 
indicates the degeneracy limit $\mbox{R}_{23}=0.75$. } }
}\vspace*{10mm}
%

Thus, we identify metallicity as the underlying cause of the observed
range in \OII/\Ha, affecting this ratio both indirectly through the
differential extinction of the \OII\ and \Ha\ lines, and directly, as
shown above.  In figure~3\emph{c} we plot the metallicity sensitive
\Rindex\ index versus absolute $B$ magnitude to explicitly show the
correlation of metallicity with luminosity. 

\null\vspace*{1.05\txw}

  
\section{Using \OII\ and \Ha\ as Tracers of the SFR}

In the following discussion we use the reddening corrected \Ha\ flux,
\Ha$_o$, to parameterize the SFR since this \Ha\ flux is directly
proportional to the ionizing flux, $F_{ion}$. 

In figure~4\emph{a} we plot the ratio of the observed \OII\ flux and the
ionizing flux (\Ha$_o$) versus the absolute $B$ magnitude.  A clear
trend with luminosity is seen:\linebreak

\noindent\leavevmode
\makebox[\txw]{
   \centerline{
      \epsfig{file=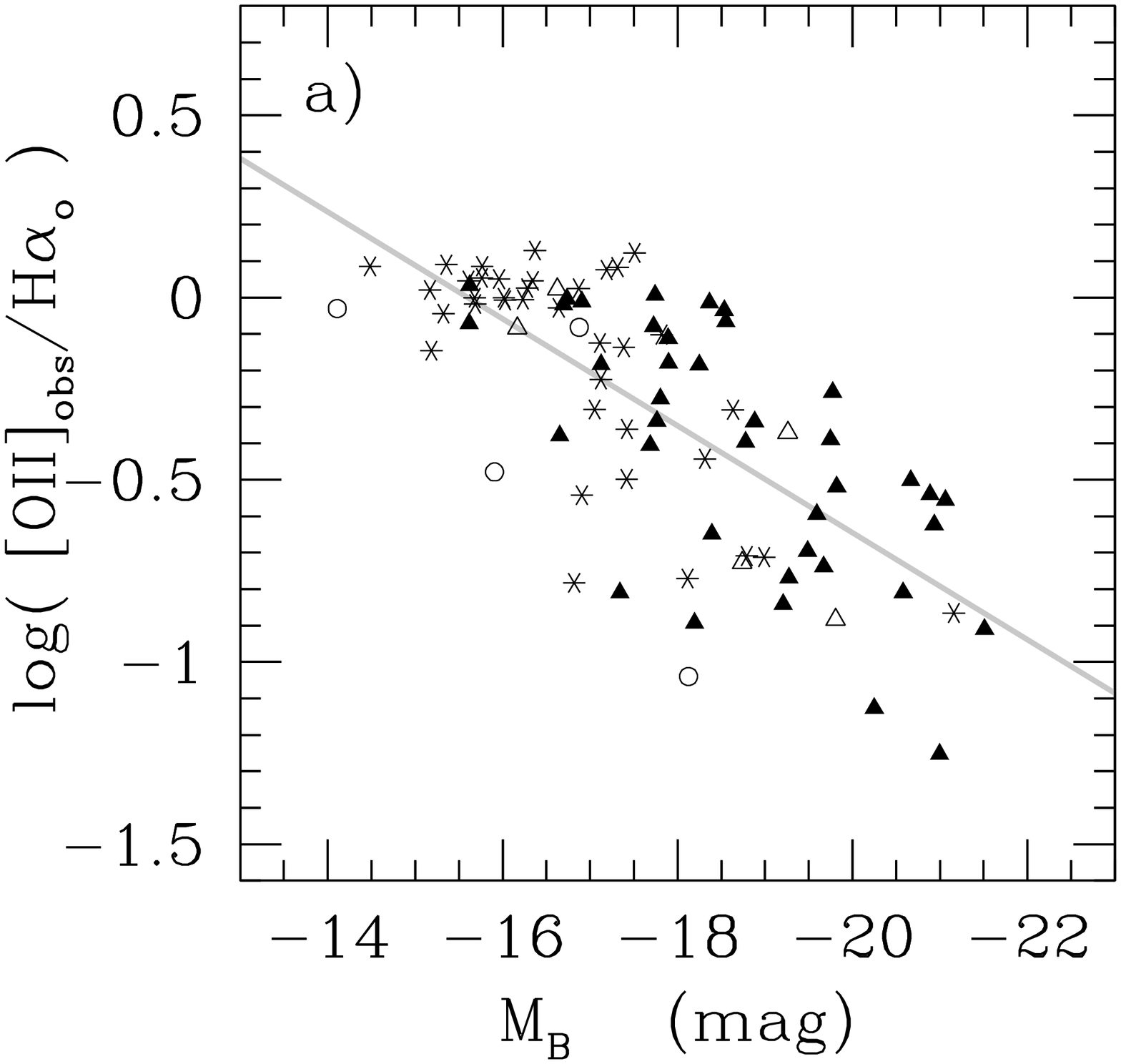,width=0.45\txw,clip=}\hspace*{0.03\txw}
      \epsfig{file=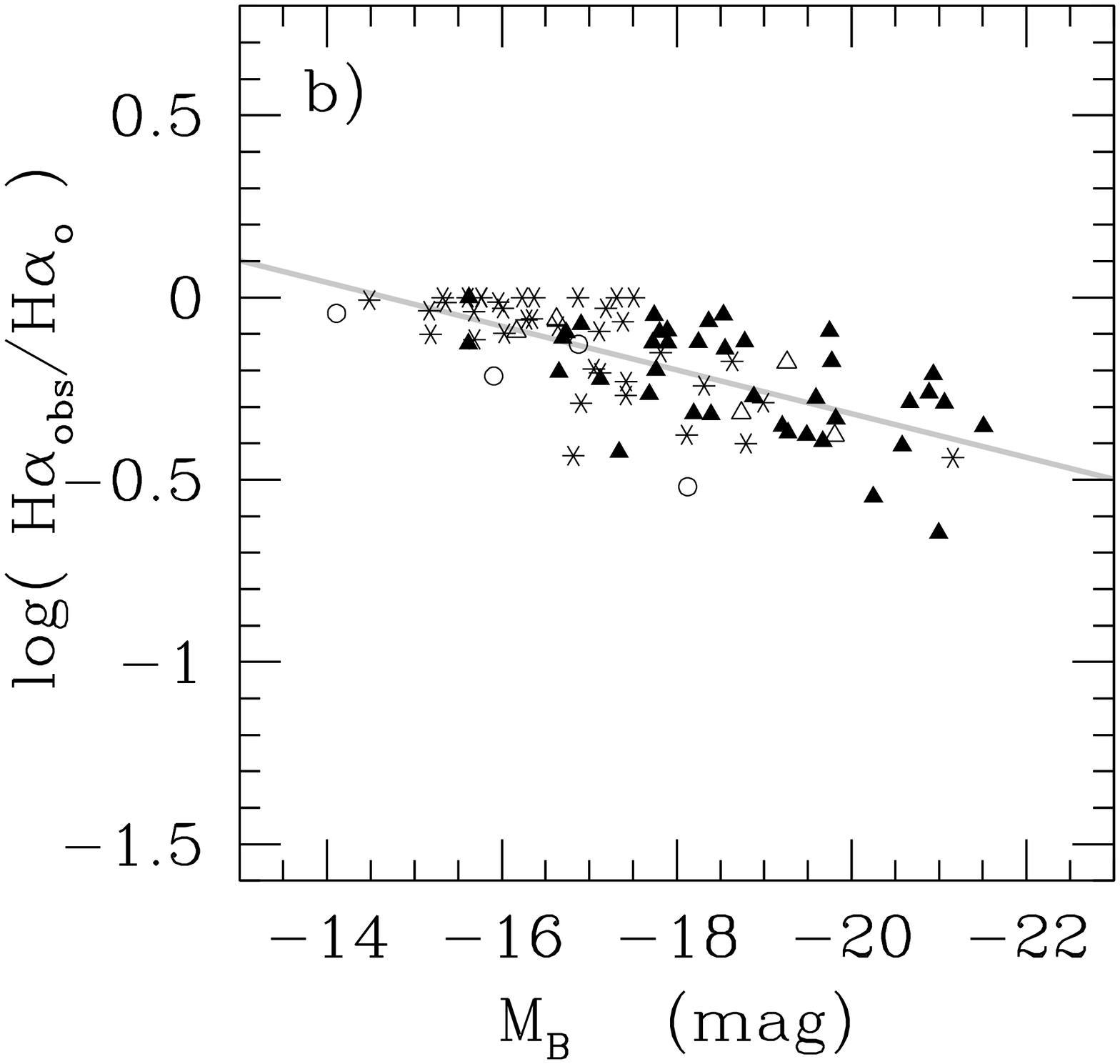,width=0.45\txw,clip=}
   }
}\par\noindent\vspace*{4mm}\makebox[\txw]{
\centerline{   
\parbox[t]{\txw}{\footnotesize {\sc Fig.~4 ---} (\emph{a}) Logarithm of
the ratio of observed \OII\ flux to ionizing flux versus total absolute
$B$ magnitude.  The zeropoint on the ordinate is arbitrary (see text). 
The plotting symbols are coded according to morphological type as in
figure~1.  The total range in $\OII_{obs}/\Ha_o$ spans 1.5 orders of
magnitude.  (\emph{b}) Logarithm of the ratio of observed \Ha\ flux to
ionizing flux versus \MB.  Imperfect correction for reddening is a much
more serious problem when using \OII\ (see figure~2\emph{a}) than when
using \Ha. } }
}\vspace*{4mm}
%

\noindent the lowest luminosity galaxies have the largest ratio of \OII\
to ionizing flux.  Near the characteristic absolute $B$ magnitude of the
local galaxy luminosity function, $M_*\sim -19$, the \OII/\Ha$_o$ ratio
varies by a factor of \tsim10.  The total range spans a factor 25.  The
SFR derived from observed \OII\ fluxes, in the absence of other
information, will be uncertain by a factor of \tsim5, if the calibration
of Kennicutt (1992) and the median absolute magnitude of his sample is
used as a reference (see section~3.1).  Using reddening corrected
instead of observed \OII\ fluxes (see figure~2\emph{b}) greatly reduces
both the luminosity dependence and the scatter at a given luminosity:
from 0.147 to 0.035 dex/mag and from 0.19 dex to 0.08 dex, respectively. 
The total range of the reddening corrected \OII/\Ha$_o$ ratio spans 0.53
dex (0.49 dex, if we exclude one outlying data point). 

As expected, reddening is a more significant source of error if \OII\
rather than \Ha\ flux is used as a SFR tracer.  In figure~4\emph{b} we
plot the ratio of the observed \Ha\ flux and the ionizing flux versus
the absolute $B$ magnitude for comparison.  The total range in the ratio
of the uncorrected and corrected \Ha\ flux spans 0.65 dex (a factor of
\tsim4.5), less than one-fifth that found for \OII.  But here as well
reddening produces a trend with galaxy luminosity of 0.060 dex/mag.

\subsection{Comparison with the Literature}

We compare our measurements for the NFGS sample with those of Kennicutt
(1992a) in order to\linebreak

\null\vspace*{0.53\txw}\noindent tie them to previous absolute
calibrations of \OII\ and \Ha\ as SFR tracers.  In figure~5 we plot
observed (\OII+\OIII\lam5007)/\Ha\ flux ratios versus observed
\OIII\lam5007/\OII\ flux ratios for both samples.  Whereas
\OIII\lam5007/\OII\ is an excitation sensitive ratio,
(\OII+\OIII\lam5007)/\Ha\ is sensitive mainly to the oxygen abundance. 
The galaxies in both samples follow similar trends.  Gallagher \etal
(1989) did not publish \OIII\lam5007 measuments for their sample, so we
are unable to make a similar comparison with their sample.  We did
however compare the observed \OII/\Hb\ flux ratios versus absolute $B$
magnitude for both the NFGS and the Gallagher \etal (1989) sample and
found them to be consistent with one another. 

Kennicutt (1992a) noted that his SFR calibration for the observed
$L(\OII)$ emission line luminosity exceeded that of Gallagher \etal
(1989) by a factor of 3.  Adopting the same IMF and conversion factor
from \Ha\ to SFR for both samples reduces this difference to a factor of
1.57 (Kennicutt 1998b).  The remaining discrepancy he ascribed to
excitation differences between the two samples. 

The median \MB\ of Kennicutt's sample is $-19.3\pm1.4$ mag, that of
Gallagher \etal is $-16.9\pm2.0$.  Taking this magnitude difference at
face value, we expect (see \S~3) a difference in the logarithm of the
\OII/\Ha$_o$ ratio of $2.4\,\hbox{mag} \cdot 0.147\,\hbox{dex/mag} =
0.353$ dex, \ie, a factor of 2.25\/.  This factor has the correct sign,
\ie, the SFR inferred from a given \OII/\Ha$_o$ ratio is higher for
Kennicutt's sample than for the sample of Gallagher \etal\ The predicted
difference\linebreak

\noindent\leavevmode
\makebox[0.475\txw]{
   \centerline{\epsfig{file=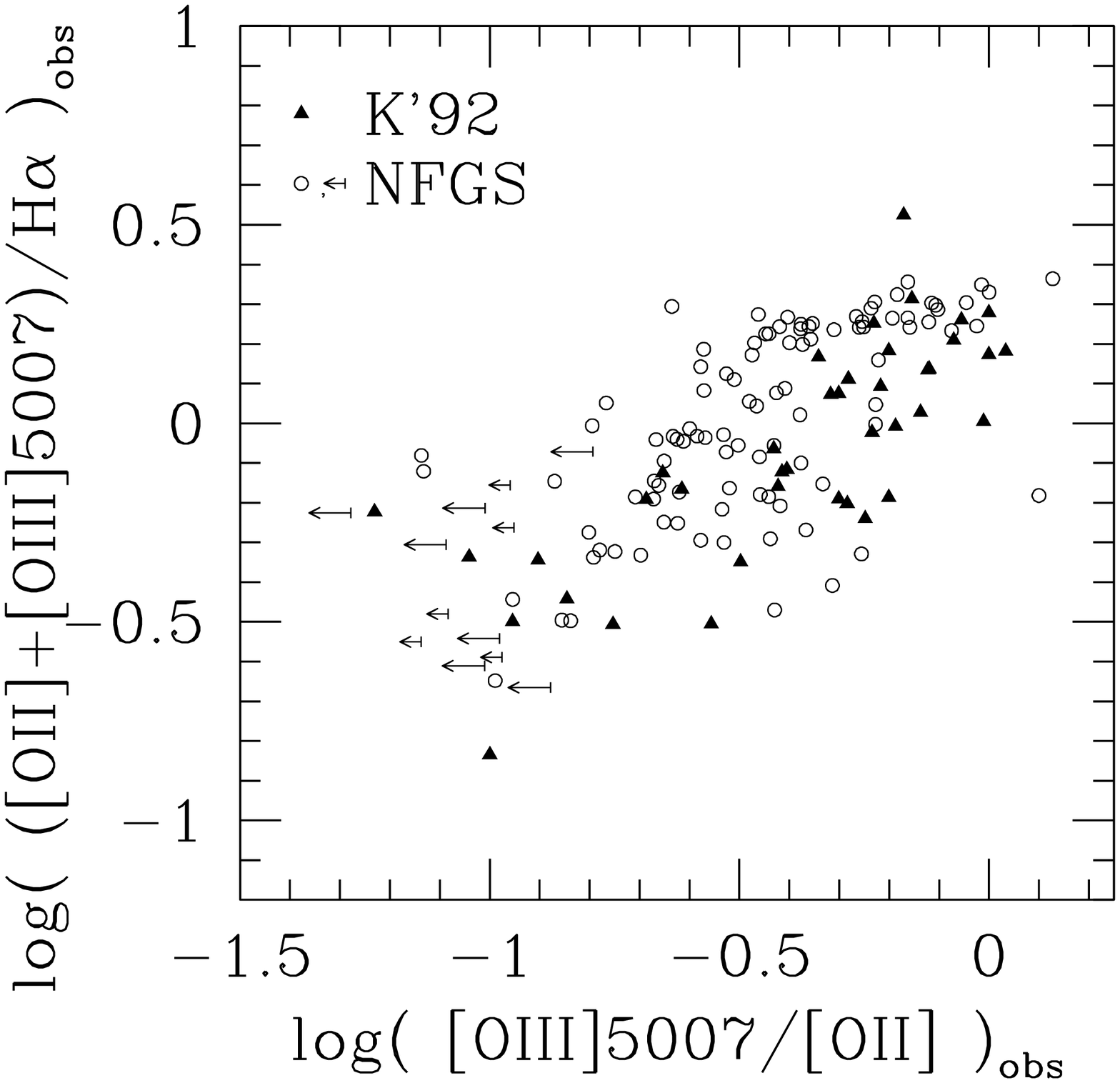,width=0.475\txw,clip=}}
}\par\vspace*{3mm}
\noindent\makebox[0.475\txw]{
\centerline{
\parbox[t]{0.475\txw}{\footnotesize {\sc Fig.~5 ---} Logarithm of the
observed (\OII+\OIII\lam5007)/\Ha\ flux ratio versus the logarithm of
the observed \OIII\lam5007/ \OII\ ratio in the sample of Kennicutt
(1992a) (labeled K'92) and in the NFGS sample (including the 118
galaxies with EW(\Ha)$<$$-10$\AA).  The abscissa is an excitation
sensitive ratio, whereas the ordinate is sensitive mainly to the oxygen
abundance.  The data points in both samples follow similar trends. 
Those data points in the present sample with errors in excess of 0.10
dex are indicated by upper limits. } }
}
%

\vspace*{7mm}

\noindent is, however, somewhat larger than that actually observed. 
This might be due to the large scatter in the trends with \MB, or to the
fact that Kennicutt could not account for the systematic variations in
absorption. 


\section{Implications for \Hb\ as a SFR Tracer}

In section~2 we showed that the observed \OII/\Ha\ ratios strongly
depend on the amount of interstellar reddening.  \Hb\ will be seriously
affected by reddening as well.  Here we test \Hb\ as a quantitative star
formation indicator. 

In figure~6\emph{a} we plot the ratio of observed \Hb\ flux and the
ionizing flux versus total absolute $B$ magnitude for the 85 galaxies
with reliable \Hb\ measurements.  The \Hb/\Ha$_o$ ratio varies
systematically with luminosity, although not as strongly as the
\OII/\Ha$_o$ ratio.  The slope of a linear fit to the data is 0.090
dex/mag and the scatter around the fit is 0.132 dex.  The \Hb/\Ha$_o$
ratio decreases from 0.37 at \MB=$-14$ to \tsim0.09 at \MB=$-21$. 

In most cases where \Hb\ can be secured, the \OIII\lam\lam4959,5007
lines will be available as well.  Figure~6\emph{b} shows the ratio of
the observed \Hb\ flux and the ionizing flux versus the observed
\OIII/\Hb\ ratio.  It is remarkable that for high values of \OIII/\Hb\
the absorption is generally very small.  The SFR can therefore be
estimated relatively well, with a conversion constant very different
from that for $L_*$ galaxies.  The scatter increases drastically for
lower values of \OIII/\Hb, and the extinction becomes much more
significant.  When no direct reddening measurement is available (\eg,
when none of the other Balmer lines can be measured reliably) the
ionizing flux can be retrieved to \tpm0.12 dex using the observed \Hb\
and \OIII\ fluxes for $\log($\OIII/\Hb$)\gtrsim0.2$.  At smaller
observed \OIII/\Hb\ ratios figure~6\emph{b} will be useful to estimate
the likely range in ionizing flux corresponding to a \Hb\
measurement.

\vspace*{2mm}


\section{Improved SFR Estimation and Empirical Corrections}

In this section our goal is to derive empirical corrections to relate
the observed \OII\ emission line flux to the ionizing flux in those
cases where \Hb\ is not available.  We start with the situation where
only \OII\ fluxes are available.  As we found above, the correlation
between absolute magnitude and \OII/\Ha$_o$ (figure~4\emph{a}) gives an
indication of whether a high or low normalization of the \OII--SFR
calibration is likely, and may be used to give a relative weight to a
set of several normalizations as published in the literature.  For
instance, at low luminosities the lower normalization of Gallagher \etal
(1989) for blue irregular galaxies is likely to be more appropriate. 
Kennicutt's (1992a) higher normalization, on the other hand, will be a
better choice for luminous galaxies.  The scatter at a given absolute
magnitude is a measure of the errors involved in the choice of
calibration constant. 

The \OII\ equivalent width, EW(\OII), can be used to estimate the likely
range in the \OII/\Ha$_o$ ratio.  In figure~7\emph{a} we plot the
logarithm of the ratio of the observed \OII\ flux and the reddening
corrected \Ha\ flux versus the logarithm of (the negative of) the
EW(\OII).  For EW(\OII)$\leq$$-45$\AA\ [\ie,
$\log\,$\{EW(\OII)\}$\geq$1.65] the scatter in \OII/\Ha$_o$ at a given
EW(\OII) is only 0.064 dex and the ionizing flux (and therefore the SFR)
can be determined well.  The \OII/\Ha$_o$ ratios for these large
EW(\OII) are high (1.04\tpm0.30), and the star formation would be
overestimated by a factor of (1.7\tpm0.5) if the SFR normalization for
luminous galaxies were used.  EW(\OII)$\leq$$-45$\AA\ are found
predominantly in the lower luminosity, lower metallicity galaxies.  For
smaller EW(\OII) we find a trend toward lower \OII/\Ha$_o$ ratios, but
the scatter in \OII/\Ha$_o$ increases to an order of magnitude.  We note
that the lack of galaxies with small EW(\OII) and large\linebreak

\noindent\leavevmode
\makebox[\txw]{
   \centerline{
      \epsfig{file=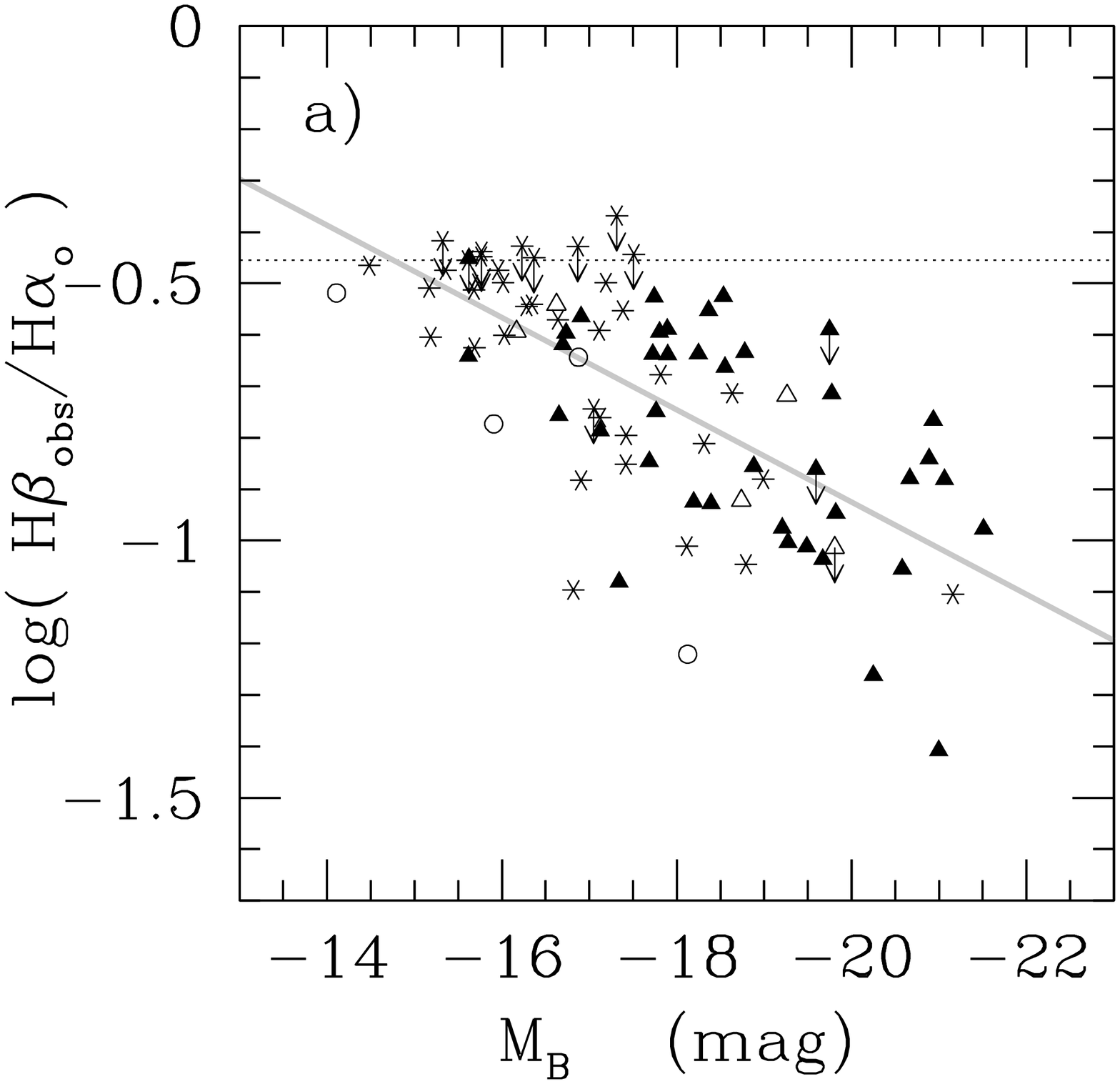,width=0.45\txw,clip=}\hspace*{0.03\txw}
      \epsfig{file=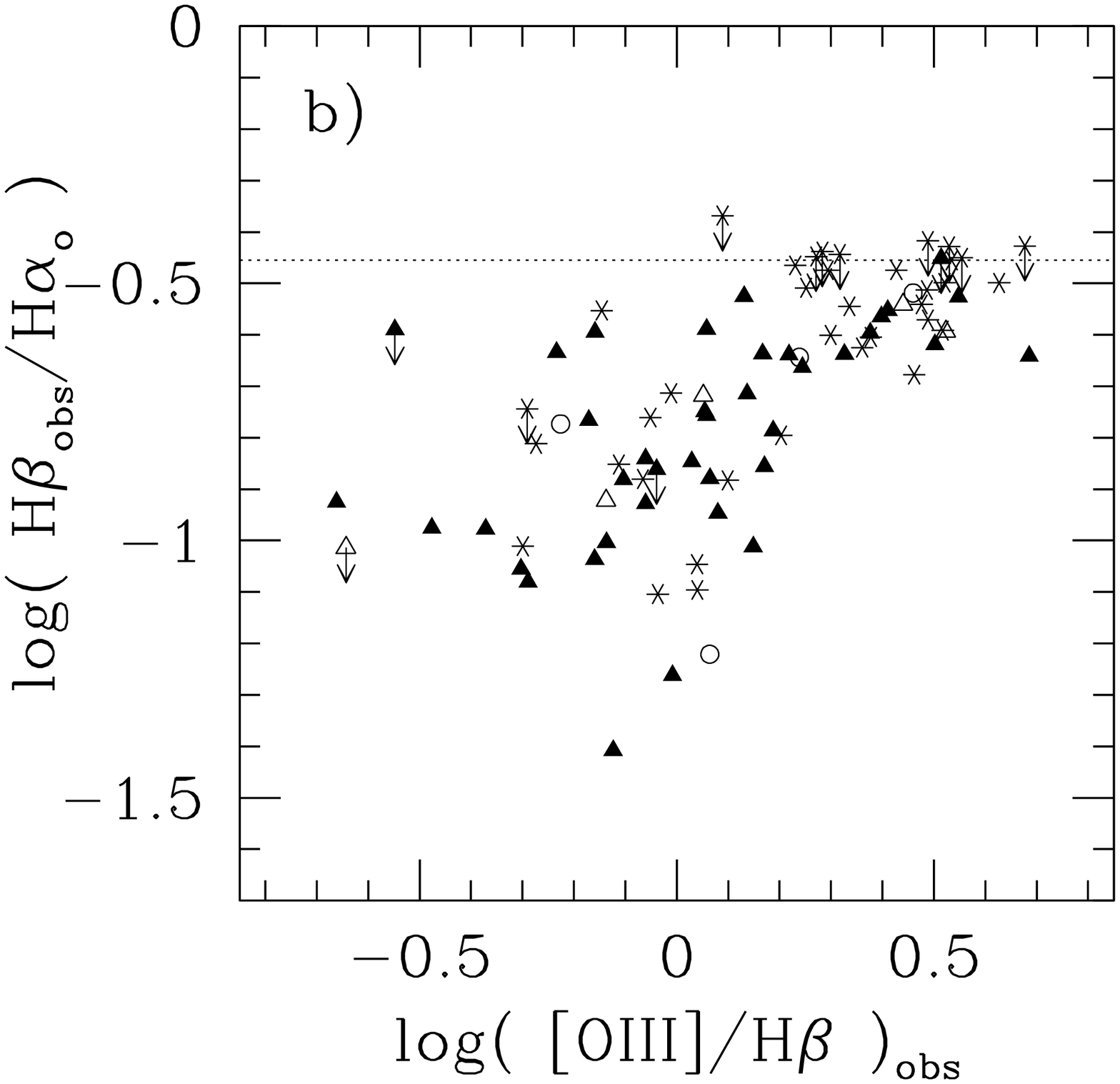,width=0.45\txw,clip=}
   }
}\par\noindent\makebox[\txw]{
\centerline{
\parbox[t]{\txw}{\footnotesize {\sc Fig.~6 ---} (\emph{a}) Logarithm of
the ratio of observed \Hb\ flux to the ionizing flux (reddening
corrected \Ha\ flux) versus absolute $B$ magnitude (including the 85
galaxies with reliable \Hb\ measurements).  The plotting symbols are
coded according to morphological type as in figure~1.  Upper limits are
used to indicate data points with errors larger than 0.15 dex.  A linear
least-squares fit to the data points is overlayed.  The dotted line
indicates the intrinsic Balmer decrement, \Ha/\Hb=2.85\/.  As was seen
for the \OII/\Ha$_o$ ratio (figure~4\emph{a}), reddening creates a
dependence on galaxy luminosity.  (\emph{b}) Logarithm of the ratio of
the observed \Hb\ flux to the ionizing flux versus the logarithm of the
observed \OIII/\Hb\ ratio.  These data can be used to estimate the
ionizing flux and its error when both \Hb\ and \OIII\ can be measured
reliably, but a direct reddening measurement is not available.  For 
$\log($\OIII/\Hb$)\gtrsim 0.2$ little reddening is seen and the ionizing
flux can be estimated accurately. } }
}\vspace*{6mm}
%

\noindent \OII/\Ha$_o$ ratios may be due to a selection effect: these
galaxies are expected to have small EW(\Hb) and will drop out of the
sample with reliable reddening corrections.  Therefore, the scatter in
\OII/\Ha$_o$ ratio for small EW(\OII) may be underestimated in
figure~7\emph{a}. 

Broadband colors help to distinguish galaxies that are particularly
dusty.  In figures~7\emph{b} and \emph{c} we plot the logarithm of the
ratio of the observed \OII\ flux to the reddening corrected \Ha\ flux
versus the effective \UB\ and \BR\ color, respectively.  Our effective
colors are the average colors measured within the half-light radius in
$B$.  Whereas the range in \OII/\Ha$_o$ for galaxies bluer than
\UBe\tsim$-0.3$ or \BRe\tsim0.95 is only \tsim0.5 dex (a scatter of
\tsim0.12 dex), redder galaxies occupy nearly the full range of observed
\OII/\Ha$_o$.  Again, the galaxies dominated by star formation have the
highest \OII/\Ha$_o$ ratios. 

The trends shown here are purely empirical, and the correlations do not
have direct physical causes.  It remains to be seen if they hold at
higher redshifts. 

\vspace*{4mm}


\section{Implications for Higher Redshifts}

Intermediate and high redshift spectroscopic studies like the Hawaii
Deep Survey (Cowie \etal 1994; Songaila \etal 1994; Cowie \etal
1996)\linebreak

\null\vspace*{0.605\txw}\noindent and the Canada-France Redshift Survey
(CFRS; Lilly \etal 1995; Hammer \etal 1997) have shown that the fraction
of relatively bright galaxies with EW(\OII)$<$$-15$\AA\ increases
strongly with redshift.  Hammer \etal (1997) showed that converting
\OII\ into a SFR using Kennicutt's (1992) calibration leads to a
production of long-lived stars in excess of the number observed at the
present epoch.  Using the Gallagher \etal (1989) calibration for blue
galaxies instead implies that 75\% of the present-day mass in stars have
been produced since $z\sim1$.  The luminosities sampled in the CFRS
survey are comparable to those in the present sample.  Hammer \etal
(1997) speculate that excitation and/or dust play a role, and our
results confirm this. 

Cowie \etal (1997) found remarkably little extinction in the bulk of the
blue star forming galaxies at redshifts $z>0.8$.  The results by Pettini
\etal (1998) for a small sample of $z\sim3$ Lyman break galaxies are
consistent with this general picture, as their UV selected galaxies
generally show low absorption compared to our $L_*$ galaxies.  The one
possible exception, DSF~2237+116~C2, is the reddest, most massive galaxy
in their sample.  These results indicate that the correlation of
reddening and excitation with luminosity may change with look-back time,
but the colors and equivalent widths might well remain effective
indicators of reddening.

\noindent\leavevmode\null\par
\makebox[\txw]{
   \centerline{
      \epsfig{file=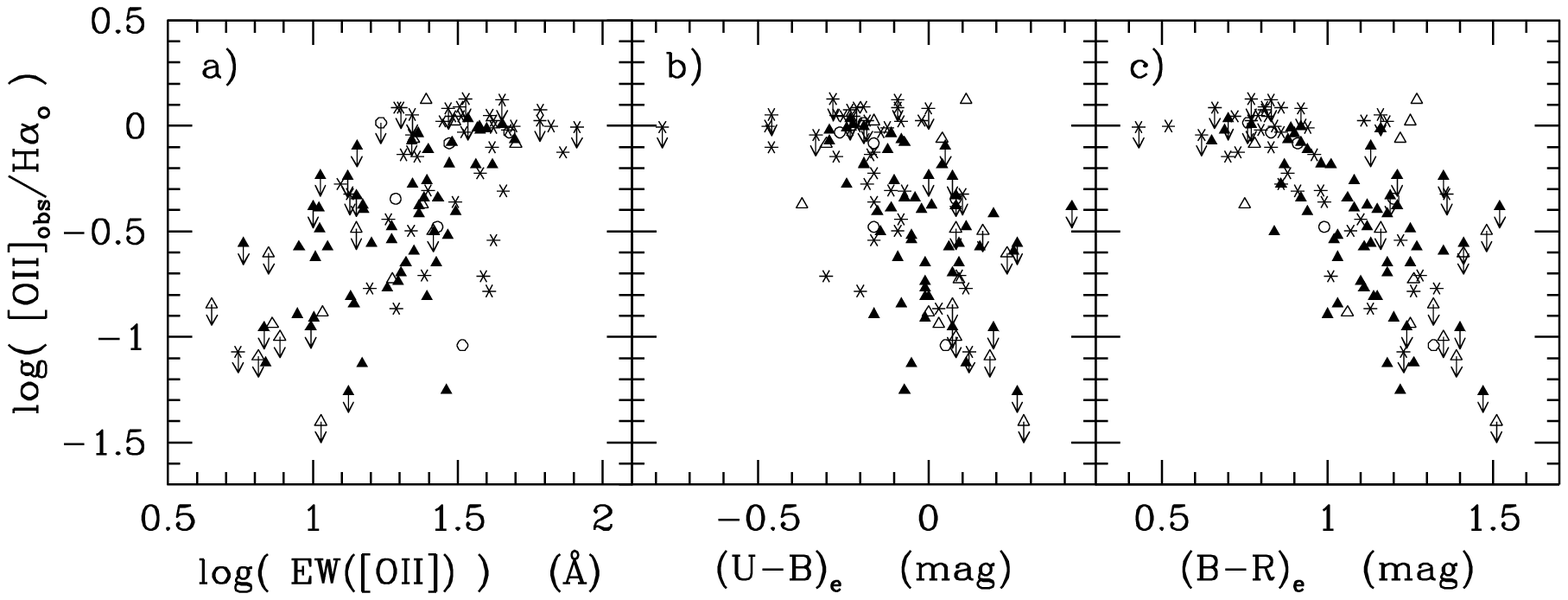,width=\txw,clip=}
   }
}\par\vspace*{4mm}\noindent\leavevmode
\makebox[\txw]{
\centerline{
\parbox[t]{\txw}{\footnotesize {\sc Fig.~7 ---} (\emph{a}) Logarithm of
the ratio of the observed \OII\ flux to the ionizing flux versus the
logarithm of (the negative of) the \OII\ equivalent width.  Plotting 
symbols are coded according to morphological type as in figure~1.  Upper
limits are used to indicate data points with errors larger than 0.15
dex.  The upper left region may be depleted due to selection effects. 
(\emph{b}) Logarithm of the ratio of the observed \OII\ flux to the
ionizing flux versus effective rest-frame \UB\ color. (\emph{c}) Same as
(\emph{b}) but using effective rest-frame \BR\ color. } }
}\vspace*{7mm}
%

The uncertainties in the SFR derived from \OII\ are of the order of
a factor of 3 if no other information is available.


\section{Summary}

We have used spectrophotometry for 85 emission line galaxies from the
Nearby Field Galaxy Survey sample to investigate the dependence of the
\OII/\Ha\ ratio on galaxy luminosity.  Reddening and excitation
differences are the main cause of the anti-correlation between \OII/\Ha\
and galaxy luminosity.  Both are strongly correlated with absolute
magnitude and are likely caused by systematic variation in metallicity
as a function of galaxy luminosity.  Excitation models match the dust
corrected line ratios within the accuracy of our data.  The total
variation in the ratio of the observed \OII\ flux to the reddening
corrected \Ha\ flux is a factor 25. 

The observed difference between the \OII--SFR calibrations of Gallagher
\etal (1989) and Kennicutt (1992a) is in the sense expected from the
difference in galaxy luminosity between the two
samples.  We confirm the conjecture of Hammer \etal (1997) that
systematic variations in reddening by dust play an important role when
interpreting emission line strengths in terms of SFRs. \newline\vfill\null

\null\vspace*{0.57\txw}\noindent When corrections for metallicity and
dust are not possible, the use of \OII\ fluxes to measure star formation
rates may result in an overestimate of a factor of 3 if local
calibrations for luminous galaxies are used.  We show that \Hb\ is a
significantly better tracer of star formation than \OII, and we discuss
some empirical trends which may be useful in the high redshift regime. 

\vspace*{4mm}


\section*{Acknowledgements}

This work was supported by grants from the University of Groningen, the
Leiden Kerkhoven-Bosscha Fund, the Netherlands Organisation for
Scientific Research (NWO), and by the Smithsonian Institution.  We thank
the CfA TAC for generously allocating time for this project over three
years.  R.\ A.\ J.\ thanks the Harvard-Smithsonian Center for
Astrophysics and the \mbox{F.\ L.~Whipple} Observatory for hospitality
during numerous visits, when all of the observations and part of this
work were carried out, and ESA's ESTEC where this work was completed. 
We thank the referee, Dr.\ J.\ S.\ Gallagher, for his thoughtful
comments that helped improve the manuscript. 

\newpage


\clearpage

\end{document}